\documentclass[useAMS,usenatbib,usegraphicx]{mn2e}

\usepackage{color}

\newcommand\apj{{ApJ}}
\newcommand\apjl{{ApJ}}
\newcommand\apjs{{ApJS}}
\newcommand\aap{{A\&A}}
\newcommand\mnras{{MNRAS}}
\newcommand\pasp{{PASP}}
\newcommand\pasj{{PASJ}}
\newcommand\qjras{{QJRAS}}
\newcommand\memsai{{Mem.~Soc.~Astron.~Italiana}}

\newcommand{\actaa}{Acta Astron.}      

\title[The helium-rich CV SBSS\,1108+574]{The helium-rich cataclysmic variable SBSS\,1108+574}
\author[P. J. Carter et al.]{P. J. Carter,$^{1}$\thanks{E-mail: philip.carter@warwick.ac.uk} D. Steeghs,$^{1}$ E. de Miguel,$^{2,3}$  W. Goff,$^{4}$ R. A. Koff,$^{5}$  T. Krajci,$^{6}$ \newauthor T. R. Marsh,$^{1}$ B. T. G\"{a}nsicke,$^{1}$ E. Breedt,$^{1}$ P. J. Groot,$^{7}$ G. Nelemans,$^{7,8}$ \newauthor G. H. A. Roelofs,$^{9}$ A. Rau,$^{10}$ D. Koester$^{11}$ and T. Kupfer$^{7}$\\
$^{1}$Department of Physics, University of Warwick, Coventry CV4 7AL\\
$^{2}$Departamento de F\'{\i}sica Aplicada, Universidad de Huelva, 21071 Huelva (Spain)\\
$^{3}$CBA (Huelva), Observatorio del CIECEM, Huelva (Spain)\\
$^{4}$CBA (Sutter Creek), 13508 Monitor Lane, Sutter Creek, CA 95685 (USA)\\
$^{5}$CBA (Colorado), Antelope Hils Observatory, 980 Antelope Drive West, Bennet, CO 80102 (USA)\\
$^{6}$CBA (New Mexico), PO Box 1351 Cloudcroft, NM 88317 (USA)\\
$^{7}$Department of Astrophysics/IMAPP, Radboud University Nijmegen, PO Box 9010, 6500 GL Nijmegen, the Netherlands\\
$^{8}$Institute for Astronomy, KU Leuven, Celestijnenlaan 200D, 3001 Leuven, Belgium\\
$^{9}$Harvard--Smithsonian Centre for Astrophysics, 60 Garden Street, Cambridge, MA 02138, USA\\
$^{10}$Max--Planck Institute for Extraterrestrial Physics, Giessenbachstr. 1, Garching 85748, Germany\\
$^{11}$Institut f\"{u}r Theoretische Physik und Astrophysik, University of Kiel, 24098 Kiel, Germany\\
}

\begin{document}

\date{Accepted 2013 January 28}

\pagerange{\pageref{firstpage}--\pageref{lastpage}} \pubyear{2013}

\maketitle

\label{firstpage}

\begin{abstract}
We present time-resolved spectroscopy and photometry of the dwarf nova SBSS\,1108+574, obtained during the 2012 outburst. Its quiescent spectrum is unusually rich in helium, showing broad, double-peaked emission lines from the accretion disc. We measure a line flux ratio He\,\textsc{i}~5875/H$\alpha$\,=\,0.81\,$\pm$\,0.04, a much higher ratio than typically observed in cataclysmic variables (CVs). The outburst spectrum shows hydrogen and helium in absorption, with weak emission of H$\alpha$ and He\,\textsc{i}~6678, as well as strong He\,\textsc{ii} emission.

From our photometry, we find the superhump period to be 56.34\,$\pm$\,0.18\,minutes, in agreement with the previously published result. The spectroscopic period, derived from the radial velocities of the emission lines, is found to be 55.3\,$\pm$\,0.8\,minutes, consistent with a previously identified photometric orbital period, and significantly below the normal CV period minimum. This indicates that the donor in SBSS\,1108+574 is highly evolved. The superhump excess derived from our photometry implies a mass ratio of $q$\,=\,0.086\,$\pm$\,0.014. Our spectroscopy reveals a grazing eclipse of the large outbursting disc. As the disc is significantly larger during outburst, it is unlikely that an eclipse will be detectable in quiescence. The relatively high accretion rate implied by the detection of outbursts, together with the large mass ratio, suggests that SBSS\,1108+574 is still evolving towards its period minimum. \end{abstract}



\section{Introduction}

Cataclysmic variable stars (CVs) consist of a white dwarf accreting from a Roche lobe filling main-sequence, slightly evolved, or brown dwarf companion. In systems where the white dwarf only has a weak or no magnetic field, an accretion disc forms around the white dwarf. See \citet{1995CAS....28.....W} for a detailed review.

From the start of mass transfer, CVs evolve towards shorter orbital periods, due to loss of orbital angular momentum via magnetic braking \citep{1981A&A...100L...7V} and gravitational wave radiation \citep{1971ApJ...170L..99F,1988QJRAS..29....1K}. A period minimum occurs when the mass of the donor becomes too low to sustain hydrogen burning, and it is driven out of thermal equilibrium, becoming partially degenerate, and no longer shrinks in response to mass loss \citep{1982ApJ...254..616R}. As the donor continues to lose mass, the orbital period increases, and the system evolves back to longer periods, with decreasing mass transfer rates. The period minimum for normal hydrogen-rich CVs is predicted theoretically to occur at $\sim$65--70\,minutes \citep{1982ApJ...254..616R,1999MNRAS.309.1034K}, and is observed at $\sim$80\,minutes \citep{2009MNRAS.397.2170G}.

There are a small number of CVs and related systems that have orbital periods below this minimum, including three confirmed CVs that have evolved donors -- stars that have been stripped of most of their hydrogen by mass-transfer or prior to the onset of mass-transfer (V485 Cen, \citealt{1996A&A...311..889A,2003ApJ...594..443G}; EI Psc, \citealt{2002ApJ...567L..49T}; CSS100603:112253$-$111037, \citealt{2012MNRAS.425.2548B}). The majority of the known ultracompact mass-transferring binaries belong to the small class known as the AM Canum Venaticorum (AM CVn) binaries. These consist of a white dwarf accreting from a hydrogen-deficient (semi-)degenerate donor, allowing them to reach their short orbital periods (5 to $\sim$65 minutes; see \citealt{2010PASP..122.1133S} for a recent review).
See \citet{2012MNRAS.425.2548B} for further discussion of the evolution and population of systems below the CV period minimum.

A subset of CVs, known as dwarf novae, shows outbursts, in which the system brightens by several magnitudes for a period of days to weeks. These outbursts are thought to be caused by thermal instabilities in the accretion disc \citep{1981A&A...104L..10M,1989PASJ...41.1005O}, and have been the subject of considerable observational and theoretical study. 
Some dwarf novae show superoutbursts in addition to the normal outbursts, these last longer and are generally brighter. During a superoutburst, tidal interactions between the disc and the donor star cause the disc to become asymmetric. This results in periodic modulations in the lightcurve, known as superhumps (e.g.\ \citealt{1988MNRAS.232...35W}). This superhump period is typically a few percent longer than the orbital period, and is related to the mass ratio of the system (e.g.\ \citealt{2005PASP..117.1204P}). The origin of superhumps is discussed in detail by \citet{2011ApJ...741..105W}.

Here we present time resolved optical spectroscopy and photometry of the helium rich CV, SBSS\,1108+574 (SDSS\,J111126.83+571238.6). The system was identified via a survey designed to uncover AM CVn binaries in the photometric database of the Sloan Digital Sky Survey \citep{2009MNRAS.394..367R,2012MNRAS.tmp..392C}. Its spectrum shows unusually strong He\,\textsc{i} emission in addition to the Balmer emission lines. It was discovered in outburst on 2012 April 22 by the Catalina Real-Time Transient Survey (CRTS; \citealt{2009ApJ...696..870D}), and reported as a new SU UMa dwarf nova by \citet{2012ATel.4112....1G}. \citet{2012arXiv1210.0678K} presented photometry of the outburst, and identified a possible orbital period of 55.367 minutes, in addition to superhumps.


\section{Observations}

\subsection{Optical spectroscopy}

%
\begin{figure*}
 \includegraphics[width=1.0\textwidth]{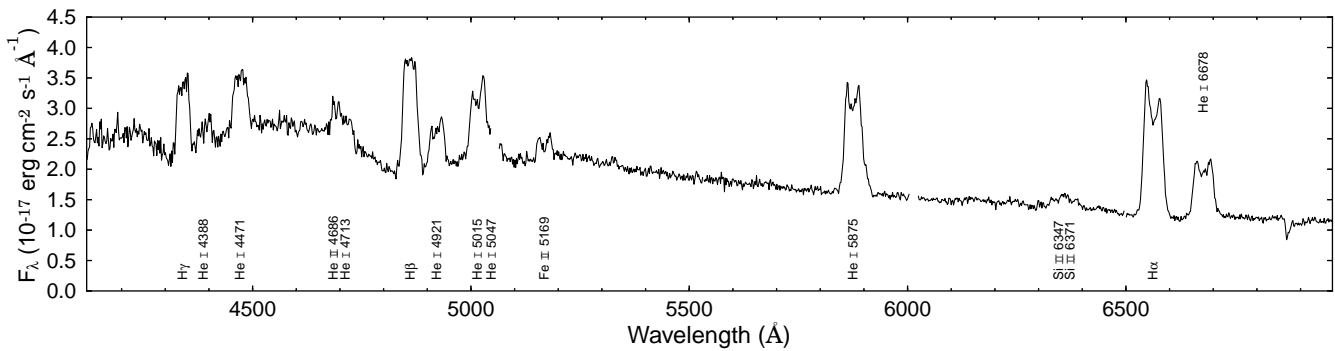}
 \caption{Average spectrum of SBSS\,1108+574 in quiescence. The most prominent lines have been labelled. \label{f:qavspec}}
\end{figure*}
We obtained time-resolved spectroscopy of SBSS\,1108+574 on 2012 February 28 and 2012 April 20 with the Gemini Multi-Object Spectrograph (GMOS; \citealt{2004PASP..116..425H}) at the Gemini-North telescope on Mauna Kea, Hawaii. We used the B600+ grating with a 1 arcsec slit. GMOS has three 2048$\times$4608 e2v deep depletion CCDs, which were used in six amplifier mode. The resulting spectra cover the wavelength range 4120\,--\,6973\,\AA{}, with an average dispersion of 1.85\,\AA{} per pixel.

The observations consist of 65 spectra in total, most of which have an exposure time of 173 seconds. The observations in February were cut short after only about one half of the binary orbit, and so are insufficient to determine the orbital period of the system. A second attempt was made to complete a 3 hour observing block on 2012 April 20, when the system was caught in an unexpected outburst (it had not been identified as a dwarf nova prior to this outburst). These spectra cover approximately three binary orbits. The log of our spectroscopic observations is given in Table \ref{t:speclog}.
\begin{table}
\centering
\caption{Summary of our spectroscopic observations of SBSS\,1108+574.}
\label{t:speclog}
\begin{tabular}{l l r r}
\hline
Date		& UT		& Exposure time (s)	& Exposures \\
\hline
2012 Feb 28	& 10:49--11:24	& 180			& 11 \\
2012 Apr 20	& 06:06--08:55	& 173			& 54 \\
\hline
\end{tabular}
\end{table}

The spectra were reduced using optimal extraction as implemented in the \textsc{Pamela}\footnote{\textsc{Pamela} is included in the \textsc{Starlink} distribution `Hawaiki' and later releases. The \textsc{Starlink} Software Group homepage can be found at http://starlink.jach.hawaii.edu/starlink.} code \citep{1989PASP..101.1032M}, which also uses the \textsc{Starlink} packages \textsc{Kappa}, \textsc{Figaro} and \textsc{Convert}. Wavelength calibration was obtained from copper-argon arc lamp exposures taken at the start of the first set of spectra, and at the start, middle and end of the observations during the second block. About 10 arc lines were identified for each of the six sections of the spectra, and fitted with fourth order polynomials, resulting in root mean square residuals of approximately 0.15\,\AA{}.

The spectra were corrected for instrumental response and flux-calibrated using the standard star Feige 34, observed as part of the Gemini baseline calibration programme. The spectra are not corrected for slit losses, so the flux calibration is not absolute, however, the relative flux calibration and continuum slope are reliable.

\subsection{Photometry}

%
\begin{table*}
\begin{minipage}{120mm}
\centering
\caption{SDSS observations of SBSS\,1108+574.}
\label{t:SDSSobs}
\begin{tabular}{l r r r r r}
\hline
MJD		& $u$	& $g$	& $r$	& $i$	& $z$ \\
\hline
52233.50	& 18.92\,$\pm$\,0.03 & 19.23\,$\pm$\,0.02 & 19.27\,$\pm$\,0.02 & 19.29\,$\pm$\,0.02 & 19.39\,$\pm$\,0.06 \\
52704.27	& 15.91\,$\pm$\,0.01 & 15.62\,$\pm$\,0.01 & 15.82\,$\pm$\,0.01 & 16.00\,$\pm$\,0.01 & 16.17\,$\pm$\,0.02 \\
52708.37	& 15.82\,$\pm$\,0.02 & 15.67\,$\pm$\,0.02 & 15.85\,$\pm$\,0.02 & 16.04\,$\pm$\,0.02 & 16.23\,$\pm$\,0.03 \\
\hline
\end{tabular}
\end{minipage}
\end{table*}

SBSS\,1108+574\footnote{SBSS\,1108+574 was assigned the CRTS ID CSS120422:111127+571239 after the outburst.} was detected in outburst on 2012 April 22 by the CRTS, at a magnitude of 15.5. This is the only outburst of the system detected by CRTS during its six year coverage. It has also been observed on three occasions by the SDSS (see Table \ref{t:SDSSobs}), indicating a previous outburst occurred on 2003 March 6, when it was detected at a $g$-magnitude of 15.6.

Differential photometry was carried out by four observers of the Centre for Backyard Astrophysics (CBA), the observations spanning a total of $\sim$135\,hours between 2012 April 23 and 2012 May 15. Details of the observers, locations and equipment are given in Table \ref{t:photinf}, and a log of our observations is given in Table \ref{t:photlog}. Unfiltered images were taken with exposure times in the range 50--120 seconds, while exposures using the $V$ filter had integration times of 120 or 240 seconds. The comparison star was either GSC 3827-0886 or GSC 3827-0824, and all data were placed onto a common magnitude scale with uncertainties no larger than $\sim$0.05 mag. Magnitudes reported here are formed from adding the $V$ magnitude of the comparison star to our differential magnitudes, and do not correspond to any standard system. Heliocentric corrections were applied to all observation times prior to analysis.
\begin{table}
\centering
\caption{Details of the observers, location, and telescope apertures used for the photometry.}
\label{t:photinf}
\begin{tabular}{l l l c}
\hline
Observer	& Code	& Location		& Aperture (inch) \\
\hline
Goff		& goff	& California, USA	& 20 \\
Koff		& koff	& Colorado, USA		& 10 \\
Krajci		& tomk	& New Mexico, USA	& 14 \\
de Miguel	& edma	& Huelva, Spain		& 16 \\
\hline
\end{tabular}
\end{table}
\begin{table*}
\begin{minipage}{108mm}
\centering
\caption{Log of our photometric observations. C corresponds to unfiltered frames.}
\label{t:photlog}
\begin{tabular}{l l l r c c}
\hline
HJD start	& HJD end	& Observer	& Exposures	& Filter	& mean magnitude \\
\hline
2456041.359	& 2456041.624	& edma		& 356		& C		& 16.05 \\
2456041.627	& 2456041.853	& koff		& 257		& C		& 16.10 \\
2456042.354	& 2456042.628	& edma		& 368		& C		& 16.21 \\
2456042.622	& 2456042.911	& koff		& 248		& C		& 16.28 \\
2456044.755	& 2456044.910	& tomk		& 265		& C		& 16.38 \\
2456045.624	& 2456045.919	& koff		& 298		& C		& 16.44 \\
2456046.660	& 2456046.938	& goff		& 170		& C		& 16.53 \\
2456047.628	& 2456047.914	& koff		& 227		& C		& 16.60 \\
2456047.660	& 2456047.958	& goff		& 96		& $V$		& 16.60 \\
2456048.411	& 2456048.461	& edma		& 48		& C		& 16.59 \\
2456052.698	& 2456052.932	& goff		& 150		& $V$		& 16.76 \\
2456053.696	& 2456053.885	& tomk		& 178		& C		& 16.78 \\
2456054.382	& 2456054.552	& edma		& 156		& C		& 16.74 \\
2456054.670	& 2456054.921	& goff		& 150		& $V$		& 16.70 \\
2456055.362	& 2456055.464	& edma		& 90		& C		& 16.65 \\
2456056.399	& 2456056.598	& edma		& 90		& C		& 16.77 \\
2456056.669	& 2456056.939	& goff		& 164		& $V$		& 16.75 \\
2456057.365	& 2456057.549	& edma		& 172		& C		& 16.69 \\
2456059.380	& 2456059.463	& edma		& 78		& C		& 16.88 \\
2456059.669	& 2456059.928	& goff		& 156		& $V$		& 16.92 \\
2456060.669	& 2456060.922	& goff		& 151		& $V$		& 16.87 \\
2456061.702	& 2456061.916	& goff		& 132		& $V$		& 16.95 \\
2456062.360	& 2456062.546	& edma		& 173		& C		& 17.07 \\
2456062.677	& 2456062.809	& goff		& 84		& $V$		& 17.14 \\
2456062.658	& 2456062.860	& koff		& 155		& C		& 17.17 \\
2456063.371	& 2456063.480	& edma		& 102		& C		& 17.15 \\
\hline
\end{tabular}
\end{minipage}
\end{table*}
%


\section{Results}

\subsection{Average spectral features}

The average quiescent spectrum of SBSS\,1108+574 is shown in Fig. \ref{f:qavspec}. It shows a blue continuum with strong broad emission lines of both hydrogen and helium. The helium lines are unusually strong, similarly to those seen in CSS1122$-$1110 \citep{2012MNRAS.425.2548B}, distinguishing SBSS\,1108+574 from other dwarf novae. Fig. 6 in \citet{2012MNRAS.tmp..392C} shows the H$\alpha$ and He\,\textsc{i} 5875 equivalent widths of this system in relation to the normal CV population. It is also worth noting the similarity between the colour of SBSS\,1108+574 and the AM CVn binaries, see fig. 11 in \citet{2012MNRAS.tmp..392C}.

The He\,\textsc{i} lines and H$\beta$ all show signs of being triple-peaked, exhibiting a central emission component in addition to the classic double-peaked emission from the accretion disc. A weak 'central spike' is a phenomenon almost never seen in CVs, but it is seen in the similar system CSS1122$-$1110. This central emission component has a low radial velocity amplitude, and is identified as originating on or close to the surface of the accreting white dwarf \citep{1999MNRAS.304..443M,2003A&A...405..249M}. It is likely that the appearance of this feature in the average spectrum (Fig. \ref{f:qavspec}) is due to the contribution of the bright spot, which does not completely average out due to incomplete coverage of the binary orbit.

\begin{figure}
 \includegraphics[width=0.48\textwidth]{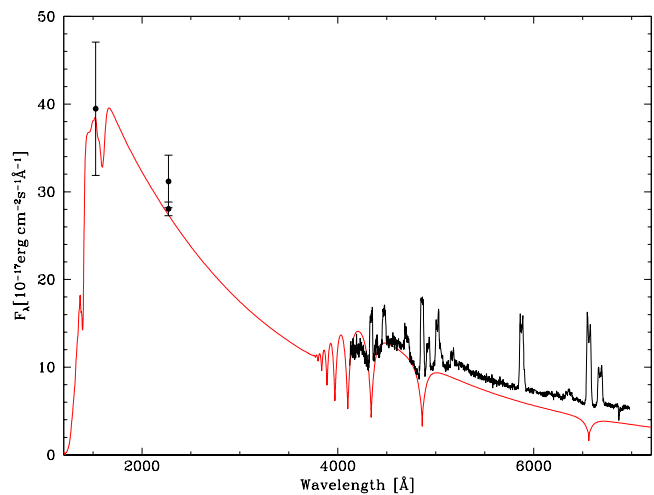}
 \caption{Shown in black are the Gemini spectrum, adjusted in flux to match the quiescent magnitude of SBSS\,1108+574 ($g=19.2$), and the UV fluxes measured by \textit{GALEX}. A white dwarf with $M_\mathrm{wd}=0.6\,\rmn{M}_\odot$ and $T_\mathrm{eff}=12000$\,K at a distance of 290\,pc (shown in red) adequately matches the broad Balmer line profiles in the GMOS spectrum, and the UV fluxes. More accurate atmospheric parameters will require far-ultraviolet spectroscopy.\label{f:wdfit}}
\end{figure}
\begin{figure*}
 \includegraphics[width=1.0\textwidth]{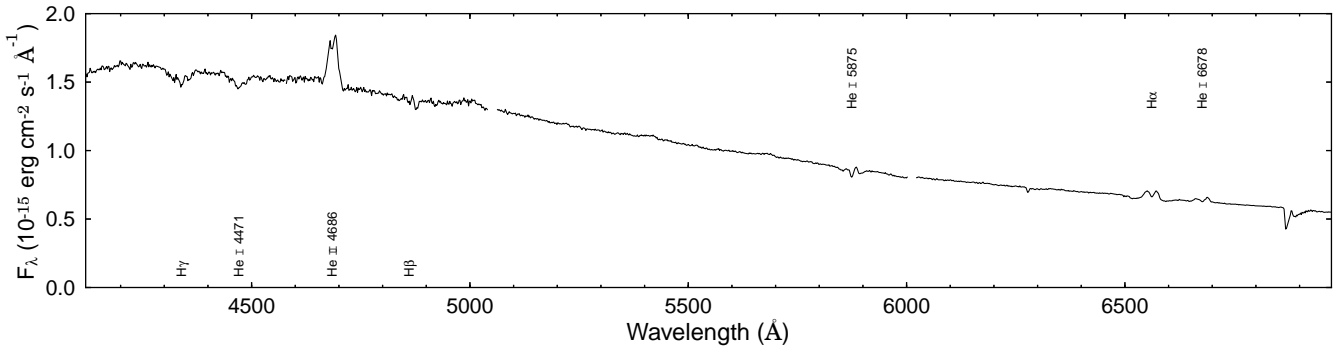}
 \caption{Average spectrum of SBSS\,1108+574 during outburst. The most prominent lines have been labelled.\label{f:avspec}}
\end{figure*}
\begin{figure*}
 \includegraphics[width=1.0\textwidth]{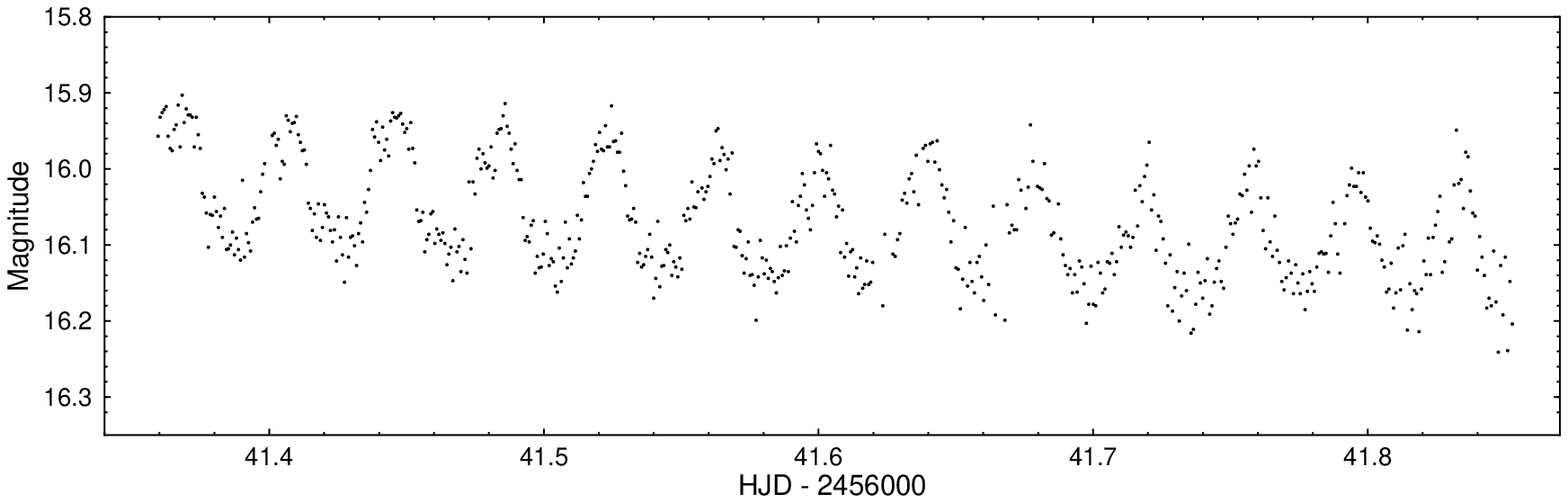}\vspace{3mm}
 \includegraphics[width=1.0\textwidth]{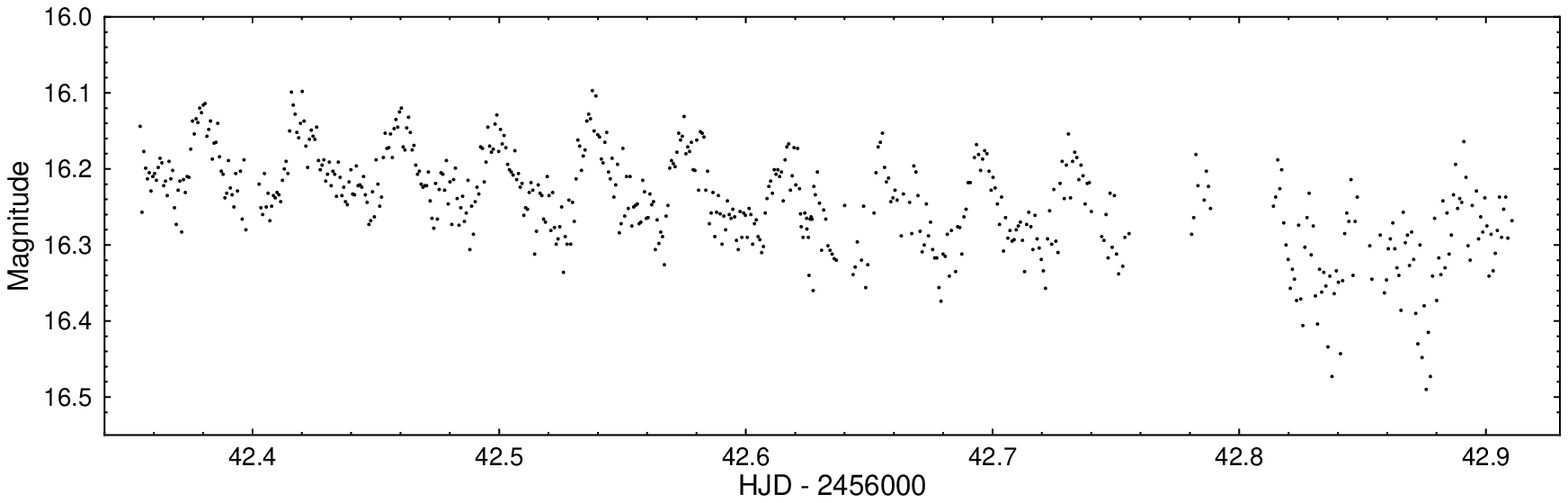}
 \caption{Lightcurves for April 23 and April 24. Each lightcurve spans $\sim$12\,hours.\label{f:nightlyLC}}
\end{figure*}

The quiescent Gemini spectrum of SBSS\,1108+574 shows broad depressions near 4340\,\AA\ and 4870\,\AA\ that we identify as H$\gamma$ and H$\beta$ absorption lines from the white dwarf photosphere. These absorption profiles are substantially perturbed by the Balmer and He emission lines, preventing a meaningful estimate of the white dwarf atmospheric properties from the optical data alone. \textit{GALEX} has detected SBSS\,1108+574 during its All-Sky Imaging Survey (AIS) in the far-ultraviolet (FUV) and near-ultraviolet (NUV) channel,
and again as part of the Medium Imaging Survey (MIS) using the NUV channel only, see Table \ref{t:galex}.
\begin{table}
\centering
\caption{\textit{GALEX} observations of SBSS\,1108+574.}
\label{t:galex}
\begin{tabular}{l l l c}
\hline
Date		& Survey	& FUV flux ($\mu$Jy)	& NUV flux ($\mu$Jy) \\
\hline
2004 Jan 25	& AIS		& $30.74 \pm 5.92$	& $53.64 \pm 5.17$ \\
2005 Jan 25	& MIS		& 			& $48.25 \pm 1.34$ \\
\hline
\end{tabular}
\end{table}
We modelled the Gemini spectrum (correcting the absolute flux to $g=19.2$, measured in quiescence by SDSS) along with the \textit{GALEX} broad-band fluxes using pure-hydrogen white dwarf atmosphere models from \citet{2010MmSAI..81..921K}. For an adopted white dwarf mass of $M_\mathrm{wd}=0.6\,\rmn{M}_\odot$ we find a temperature of $T_\mathrm{eff}=12000\pm1000$\,K, and a distance of $d=290\pm30$\,pc (Fig. \ref{f:wdfit}). Allowing for $\pm0.2\,\rmn{M}_\odot$ in the white dwarf mass adds another $\simeq\pm1000$\,K in the uncertainty of the temperature.

The vast majority of short-period CVs are relatively old systems with typical ages of several Gyr (e.g.\ \citealt{1998MNRAS.298L..29K,2003A&A...406..305S,2011ApJS..194...28K}). The cooling ages implied by their white dwarf primaries are generally much shorter~--~in the case of SBSS\,1108+574, the effective temperature corresponds to $\tau_\mathrm{cool}\simeq4\times10^8$\,yr only. The solution to this conundrum is that the white dwarfs are (re-)heated by accretion. Short-term fluctuations of the mass transfer rate, such as dwarf nova outbursts, affect only their outer envelope of a white dwarf, and the effective temperature responds on time scales of weeks to years (e.g.\ \citealt{1996A&A...309L..47G,1998ApJ...496..449S,1999PASP..111.1292S}). In contrast, the effective temperature determined sufficiently long after a dwarf nova outburst reflects the secular mean of the accretion rate over $\sim10^5$\,yr (for low $\dot M$ systems, \citealt{2003ApJ...596L.227T,2009ApJ...693.1007T}). Adopting a white dwarf mass of $M_\mathrm{wd}=0.6\,\rmn{M}_\odot$ ($M_\mathrm{wd}=0.9\,\rmn{M}_\odot$), we estimate $\dot M\sim10^{-10}\,\rmn{M}_\odot\,\mathrm{yr}^{-1}$($\dot M\sim3\times10^{-11}\,\rmn{M}_\odot\,\mathrm{yr}^{-1}$).

In addition to the hydrogen and helium lines, Fe\,\textsc{ii} emission at 5169\,\AA{}, and Si emission at 6347\,\AA{} and 6371\,\AA{} are identified.

The average outburst spectrum is shown in Fig. \ref{f:avspec}. It shows a bluer continuum than the quiescent spectrum, and both helium and hydrogen in absorption, a common feature in outbursting dwarf novae due to the bright, optically thick disc. It also shows He\,\textsc{ii} 4686, H$\alpha$ and He\,\textsc{i} 6678 in emission, with the classic double-peaked profile caused by the accretion disc.

Table \ref{t:ew} shows the reduction in equivalent widths of the prominent lines between quiescence and outburst. The ratio of the equivalent widths of H$\alpha$ to He\,\textsc{i} 5875 is approximately 3 times smaller than typically found in CVs \citep{2012MNRAS.tmp..392C}. From the quiescent spectrum we calculate a line flux ratio He\,\textsc{i}~5875/H$\alpha$ = 0.81\,$\pm$\,0.04, a much higher ratio than normally observed in cataclysmic variables, where it is typically 0.2 -- 0.4 (e.g.\ \citealt{1982ApJ...257..672W,2001MNRAS.326.1235T,2012MNRAS.425.2548B}).
\begin{table*}
\centering
\begin{minipage}{110mm}
\caption{Equivalent widths and FWHM of prominent lines. Estimated uncertainties are for a 1$\sigma$ confidence level.}
\label{t:ew}
\begin{tabular}{l l l r l}
\hline
Line			& \multicolumn{2}{c}{Quiescent} 		& \multicolumn{2}{c}{Outburst} \footnotetext{\hspace{-2mm}Values marked '...' could not be measured reliably.} \\
			& EW (\AA{})		& FWHM (km s$^{-1}$)	& EW (\AA{})		& FWHM (km s$^{-1}$) \\
\hline
H$\gamma$		& $-$17.4\,$\pm$\,2.0	& 2100\,$\pm$\,50		& 3.7\,$\pm$\,0.3		& 3500\,$\pm$\,100 \\
He\,\textsc{i} 4471	& $-$12.9\,$\pm$\,1.5	& 2200\,$\pm$\,50		& 1.5\,$\pm$\,0.2		& 2000\,$\pm$\,100 \\
He\,\textsc{ii} 4686	& $-$13.7\,$\pm$\,1.5\footnote{Blended with He\,\textsc{i} 4713.}	& 3300\,$\pm$\,200 & $-$6.9\,$\pm$\,0.1	& 1600\,$\pm$\,50 \\
H$\beta$		& $-$32.0\,$\pm$\,2.0	& 1900\,$\pm$\,50		& 1.6\,$\pm$\,0.3		& 4400\,$\pm$\,300 \\
He\,\textsc{i} 4921	& $-$11.0\,$\pm$\,1.5	& 1900\,$\pm$\,100	& ... 			& ... \\
He\,\textsc{i} 5015\footnote{The red wing falls in the gap between the CCDs, and may be blended with He\,\textsc{i} 5047.} & $-$18.8\,$\pm$\,0.5 & 2400\,$\pm$\,100 & ... & ... \\
He\,\textsc{i} 5875	& $-$49.0\,$\pm$\,1.0	& 2100\,$\pm$\,50		& 1.3\,$\pm$\,0.5		& 2300\,$\pm$\,100 \\
H$\alpha$		& $-$84.0\,$\pm$\,3.0	& 1950\,$\pm$\,50		& $-$5.0\,$\pm$\,1.0	& 1700\,$\pm$\,50 \\
He\,\textsc{i} 6678	& $-$35.5\,$\pm$\,1.4	& 2000\,$\pm$\,100	& $-$1.3\,$\pm$\,0.5	& 1800\,$\pm$\,200 \\
\hline
\vspace{-1cm}
\end{tabular}
\end{minipage}
\end{table*}
%

\subsection{Photometric periods}

Nightly lightcurves for April 23 and April 24 are shown in Fig. \ref{f:nightlyLC}. A periodic signal of amplitude $\sim$80\,mmag is clearly seen during the first two nights.

\begin{figure}
 \includegraphics[width=0.48\textwidth]{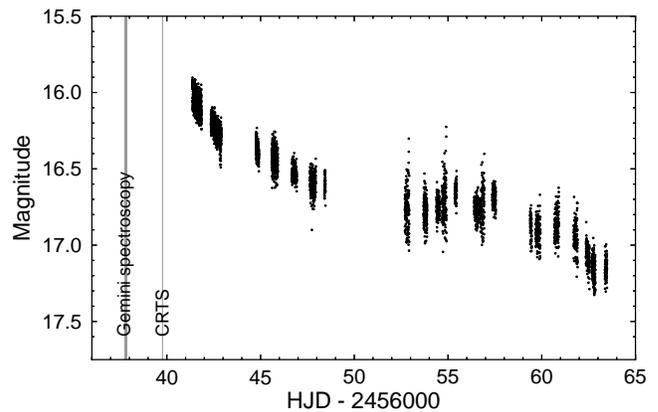}
 \caption{Full lightcurve of our photometric observations, showing the times of our spectroscopic observations and the detection by the CRTS.\label{f:fullLC}}
\end{figure}
The full lightcurve of our observations is shown in Fig. \ref{f:fullLC}. The system fades at a nearly constant rate of $\sim$0.10\,mag\,d$^{-1}$ over the first week. This is followed by a period of $\sim$6 -- 10\,days over which the mean magnitude remains essentially constant; afterwards, the system resumes its way to its low state. Oscillations were observed in all nightly light curves, although they tended to become less well-defined as the system turned fainter and the signal-to-noise ratio decreased. The amplitude was not seen to change appreciably.

\begin{figure}
 \includegraphics[width=0.48\textwidth]{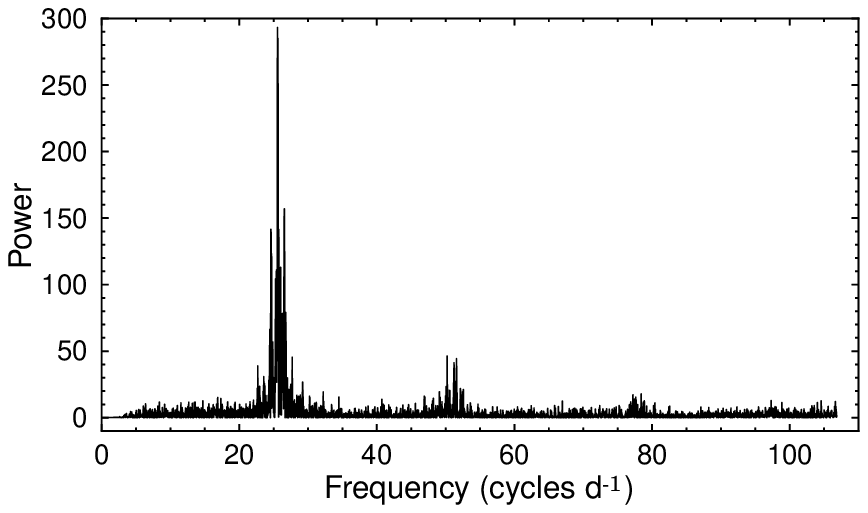}\vspace{3mm}
 \includegraphics[width=0.48\textwidth]{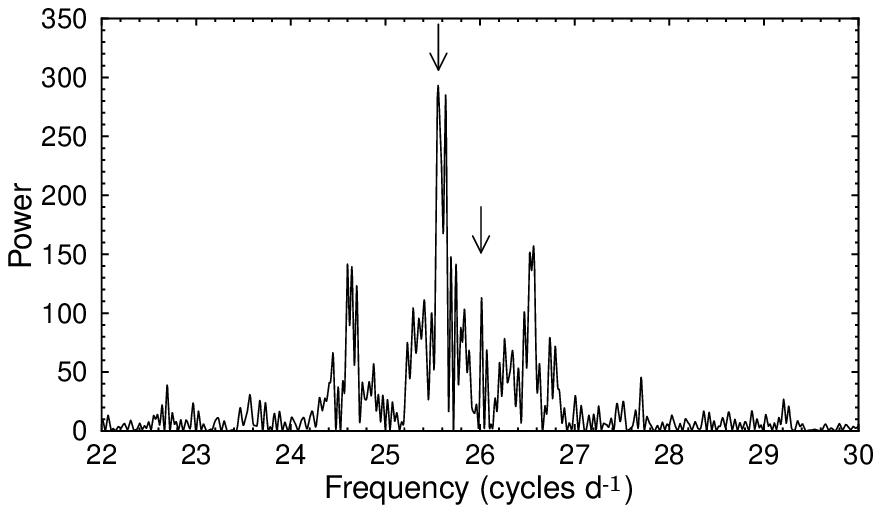}
 \caption{Lomb-Scargle periodogram of the photometric data set. The lower panel shows the detail around the strongest peak. The fundamental frequency and an additional peak that is not identified as due to aliasing, are indicated by arrows. We interpret these signals as the superhump period and orbital period of the binary.\label{f:pgram}}
\end{figure}
The Lomb-Scargle periodogram of the whole data set is shown in Fig. \ref{f:pgram}. We first removed any obvious trend in the nightly light curves and subtracted the nightly mean magnitude. The frequency of the fundamental signal is found at 25.56\,cycles\,d$^{-1}$, corresponding to a period of 56.34\,minutes. The presence of a close peak at 25.64\,cycles\,d$^{-1}$ with nearly the same power is an indication of variations in the periodicity of the fundamental signal during the time spanned by our observations. As the superhump signal is not totally periodic, the power spreads out, resulting in the second peak. This period instability is typical of superhumps and rules out a possible orbital nature of the fundamental signal. The $\pm$1\,cycle\,d$^{-1}$ aliases are also clear in Fig. \ref{f:pgram}.

\begin{figure}
 \includegraphics[width=0.48\textwidth]{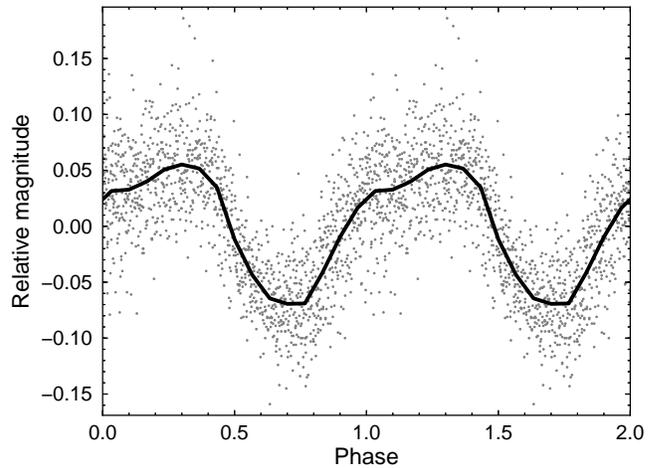}
 \caption{Lightcurve of the first two nights of observations, folded on the detected superhump period, 56.34\,minutes. The zero phase is arbitrary.\label{f:foldedLC}}
\end{figure}

An $O\,-\,C$ analysis of our data confirms the variation in the period over the time spanned by our observations. This period instability has a significant impact on the accuracy with which we can determine the superhump period, we therefore take the strongest peak in the periodogram as our superhump period, and use the separation of the two strong peaks as an estimate of the uncertainty on this period, 56.34\,$\pm$\,0.18\,minutes. This estimated uncertainty is consistent with the variation revealed by the $O\,-\,C$ analysis. The phase-folded lightcurve constructed from the first two nights of observations is shown in Fig. \ref{f:foldedLC}.
\begin{figure}
 \includegraphics[width=0.48\textwidth]{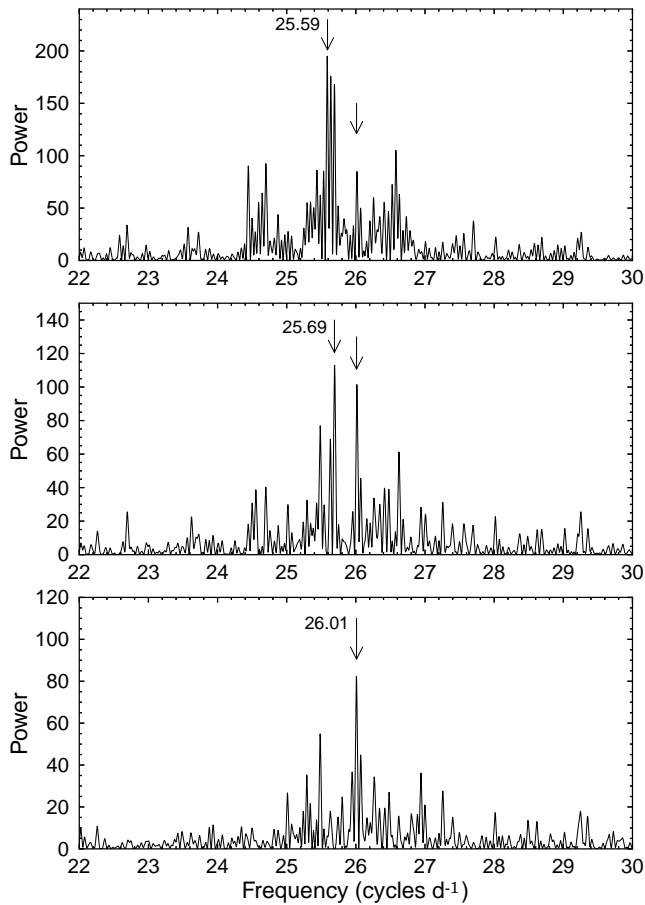}
\caption{Periodograms of the photometric data set after prewhitening. The top panel shows the periodogram after prewhitening the data with the previously identified strongest frequency at 25.56\,cycles\,d$^{-1}$ (see Fig. \ref{f:pgram}). The strongest peak is at 25.59\,cycles\,d$^{-1}$. The middle panel shows the periodogram after prewhitening the previous lightcurve with the 25.59\,cycles\,d$^{-1}$ signal. The strongest peak is now found at a frequency of 25.69\,cycles\,d$^{-1}$. The lower panel shows the periodogram after prewhitening the previous lightcurve with the 25.69\,cycles\,d$^{-1}$ signal. The strongest peak is now found at 26.01\,cycles\,d$^{-1}$, and is identified as corresponding to the orbital motion. The strongest peak and the 26.01\,cycles\,d$^{-1}$ signal are indicated by arrows in each panel.\label{f:prewhite}}
\end{figure}
There is an additional peak centred at 26.01\,cycles\,d$^{-1}$ which seems to bear no relation with the time windowing of our observations. This is identified as corresponding to the underlying orbital motion, giving a period of 55.36\,$\pm$\,0.03\,minutes, and confirming the findings of \citet{2012arXiv1210.0678K}. This signal remains after prewhitening the light curve with the main frequency at 25.56\,cycles\,d$^{-1}$, and is the strongest signal remaining after sequentially prewhitening the residual light curve with two further strong frequencies (25.59 and 25.69\,cycles\,d$^{-1}$) associated with the superhumps (see Fig. \ref{f:prewhite}). We also find relatively strong signals corresponding to the third and fourth harmonics of this frequency. The strength of these peaks would not be unexpected given the sharp dip that would be caused by a possible eclipse of the disc (see section \ref{s:eclipse}). This supports the identification of this frequency as corresponding to the orbital period.

\subsection{The spectroscopic period}

To find the spectroscopic period we measured the radial velocity variation of the emission lines using the single Gaussian technique of \citet{1980ApJ...238..946S}, as implemented in \textsc{Molly}\footnote{\textsc{Molly} was written by T. R. Marsh and is available from http://www.warwick.ac.uk/go/trmarsh/software.}. We varied the full width half maximum of this Gaussian from 200 to 3000\,km\,s$^{-1}$. For each resulting radial velocity curve we calculated the Lomb-Scargle periodogram \citep{1982ApJ...263..835S}, and fit the radial velocities with a circular orbit of the form
\begin{equation}\label{e:rv}
V(t) \, = \, K \; \rmn{sin}\left(2\pi(t-t_0) \over P_{\rmn{orb}} \right) \; + \; \gamma,
\end{equation}
where $P_{\rmn{orb}}$ is derived from the strongest peak in the periodogram.

\begin{figure}
 \includegraphics[width=0.48\textwidth]{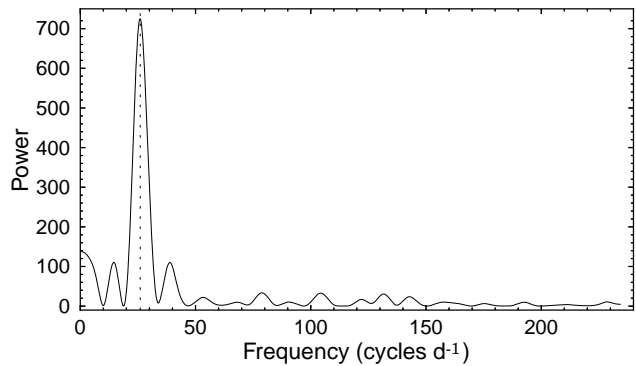}
 \caption{Lomb-Scargle periodogram calculated from the He\,\textsc{ii} 4686 radial velocities of SBSS\,1108+574. The location of the orbital period identified from the photometry is indicated by the dotted line for comparison. \label{f:4686pgram}}
\end{figure}

Examining the results, we determine that the errors are smallest for a FWHM of $\sim$2200\,km\,s$^{-1}$ for the He\,\textsc{ii} 4686 line. We use such a wide Gaussian as this gives an average of the velocity over the entire line, minimising the errors introduced by the unusual structure of the lines from the outbursting disc. We show the Lomb-Scargle periodogram calculated from the He\,\textsc{ii} radial velocities in Fig. \ref{f:4686pgram}. A clear signal is seen at 26.03\,cycle\,d$^{-1}$, corresponding to a period of 55.3\,minutes.
\begin{figure}
 \includegraphics[width=0.48\textwidth]{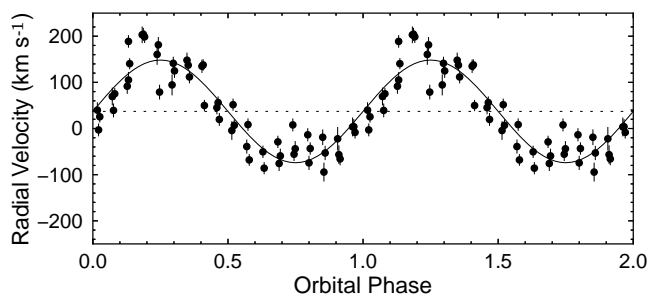}
 \caption{Measured He\,\textsc{ii} radial velocities folded on a period of 55.3\,minutes. The solid and dotted lines are the best fit radial velocity curve and $\gamma$ velocity, the parameters are shown in Table \ref{t:RVfits}. \label{f:4686rv}}
\end{figure}
The He\,\textsc{ii} 4686 radial velocity curve, folded on this period, is shown in Fig. \ref{f:4686rv}. The parameters of the fits to Equation \ref{e:rv}, derived from the radial velocity measurements, are shown in Table \ref{t:RVfits}. We take the zero phase, HJD$_0$, as being defined by the blue to red crossing of the velocities.
\begin{table*}
\begin{minipage}{100mm}
\centering
\caption{Orbit parameters derived from radial velocity measurements}
\label{t:RVfits}
\begin{tabular}{c c c c}
\hline
$P_{\rmn{orb}}$ (min)		& HJD$_0$		& $K$ (km\,s$^{-1}$)	& $\gamma$ (km\,s$^{-1}$) \\
\hline
55.3\,$\pm$\,0.8		& 2456037.7551(4)	& 111\,$\pm$\,7		& 37.1\,$\pm$\,5.0 \\
\hline\end{tabular}
\end{minipage}
\end{table*}

There is considerable scatter around the radial velocity curve shown in Fig. \ref{f:4686rv}, which is likely caused by the unusual line structure during the outburst (see Fig. \ref{f:trail}). Calculating the radial velocities from the line wings using two narrower Gaussians gives similar results, but the effect of the unusual structure causes greater scatter in the radial velocities, and greater errors, and so we prefer the single Gaussian method.

The uncertainty on the period was estimated by carrying out 10\,000 bootstrap selections of the radial velocity curve. For each subset, 54 radial velocities were selected from the full radial velocity curve, allowing for points to be selected more than once, and the periodogram calculated, taking the strongest peak as the period. The standard deviation of these computed periods, ignoring those that correspond to higher harmonics, is taken as a measure of the uncertainty in the derived orbital period.

\subsection{Mass ratio}

Superhumps are caused by a resonant interaction between the accretion disc and the donor star, that causes the disc to become asymmetric. The increased viscous dissipation caused by this interaction between the donor and the distorted disc, leads to the brightness variations observed during dwarf nova superoutbursts. The observed superhump period is the beat period between the orbital period of the system and the precession period of the deformed disc \citep{1988MNRAS.232...35W}.

As the superhump phenomenon is due to resonance, and the precession rate for resonant orbits depends on the mass ratio of the system, there is a strong link between the superhump period and the mass ratio. The superhump period-excess,
\begin{equation}\label{e:shexcess}
\epsilon = (P_{\rmn{sh}} - P_{\rmn{orb}}) / P_{\rmn{orb}},
\end{equation}
is found to increase with increasing mass ratio, $q$. An empirical relation is derived from eclipsing dwarf novae, in which the mass ratio and superhump excess can be measured independently \citep{2005PASP..117.1204P,2006MNRAS.373..484K,2009PASJ...61S.395K}.

Of the three alternative $\epsilon$ -- $q$ relations, the \citet{2005PASP..117.1204P} form,
\begin{equation}\label{e:e-q}
\epsilon = 0.18q + 0.29q^2,
\end{equation}
and the \citet{2009PASJ...61S.395K} form, both assume $\epsilon = 0$ when $q = 0$. This is a reasonable assumption as we would expect a secondary with negligible mass to have a negligible tidal interaction with the disc. The third version of the relation, given by \citet{2006MNRAS.373..484K}, does not use this assumption. The \citet{2005PASP..117.1204P} formulation is usually favoured for AM CVn binaries \citep{2012MNRAS.425.2548B}, as observations of the only known eclipsing system, SDSS J0926+3624 agree best with this form \citep{2011MNRAS.410.1113C}, and so we use that version here.

The total time covered by our spectroscopic observations is not sufficiently long to reach the accuracy of the photometric periods, and so we take the weak additional signal in Fig. \ref{f:pgram} as the orbital period. As the superhump period varies during our observations, we conservatively take the separation of the two strongest peaks in Fig. \ref{f:pgram} as a measure of the uncertainty on the superhump period for calculation of the excess. This gives the superhump excess in SBSS\,1108+574, $\epsilon$\,=\,0.0176\,$\pm$\,0.0032, and the mass ratio, $q$\,=\,0.086\,$\pm$\,0.014.
We note that this is very large compared to the mass ratio found for CSS1122$-$1110, $q$\,=\,0.017 \citep{2012MNRAS.425.2548B}, despite the similar orbital periods.

Our larger estimate for the superhump excess is consistent with the value found by \citet{2012arXiv1210.0678K} for their stage B superhumps, $\epsilon$\,=\,0.0174\,$\pm$\,0.0002. Using the smaller excess they measured for their stage C superhumps, and the \citet{2009PASJ...61S.395K} form of the $\epsilon$ -- $q$ relation, they derive a mass ratio $q$\,=\,0.06. We note that the values we derive for $q$ using the \citet{2005PASP..117.1204P} form of the  $\epsilon$ -- $q$ relation, are consistent with the values calculated using the \citet{2006MNRAS.373..484K} form. For our data the \citet{2009PASJ...61S.395K} formulation gives a larger value of $q$, however, this relation is calibrated using the shorter superhump periods that occur late in the superoutburst, and so likely overestimates $q$ for our longer periods from earlier in the superoutburst.

\subsection{Dynamic spectrum}

%
\begin{figure*}
 \includegraphics[width=1.0\textwidth]{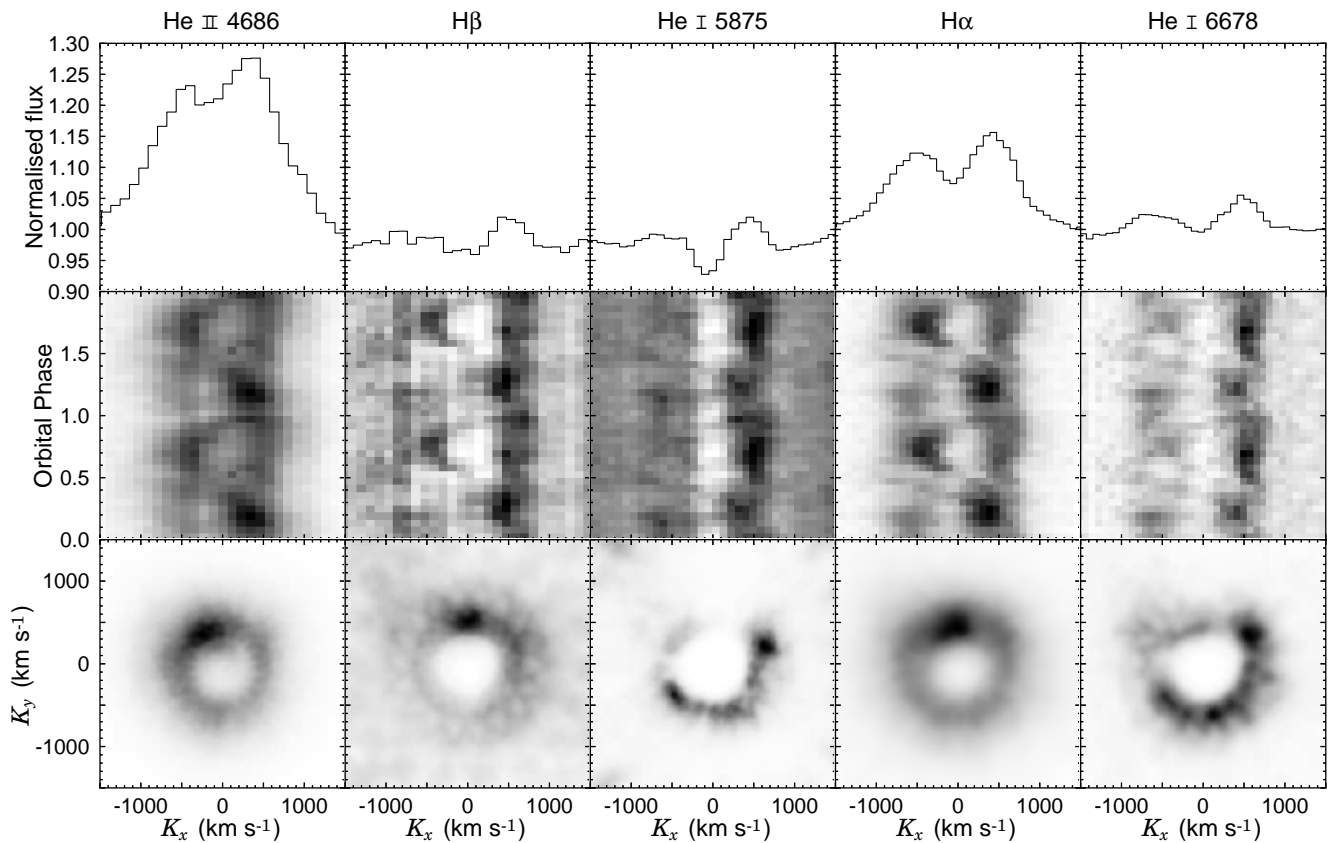}
 \caption{Average continuum-normalised line profiles, phase-folded trailed spectra, and Doppler maps of the strongest lines in the outburst spectra. The tomograms for H$\beta$ and He\,\textsc{i} 5875 were created after subtracting a spline fit to the background.\label{f:trail}}
\end{figure*}
The line profiles and trailed spectra for the strongest lines, folded on the orbital period, 55.3\,minutes (26.03\,cycles\,d$^{-1}$), are shown in Fig. \ref{f:trail}. It is clear that there is greater flux in the redshifted line peak than the blueshifted peak; this unexplained asymmetry is often seen in the spectra of outbursting CVs.
We also note the presence of underlying absorption due to the optically thick outbursting disc, this is particularly noticeable in the He\,\textsc{i} 5875 line, and makes the S-wave difficult to discern when it crosses the line centres.

An S-wave is clear in the strongest line, He\,\textsc{ii} 4686, but becomes weaker with decreasing line strength. The presence of this S-wave can be seen in the other lines shown in Fig. \ref{f:trail}, however, the He\,\textsc{i} lines appear to show a brighter varying signal with higher velocities than the S-wave, that is almost in anti-phase. A second varying signal is also seen in the H$\alpha$ trailed spectrum. This feature is likely responsible for the strength of the second harmonic in the periodograms of these lines. There is no coherent S-wave visible in the trailed spectra when folded on these second harmonic frequencies; thus we are confident that we have identified the correct period of this system.

The corresponding Doppler tomograms \citep{1988MNRAS.235..269M} give a similar picture of emission in the disc. A bright spot is clearly visible in the maps for He\,\textsc{ii} 4686, H$\alpha$ and H$\beta$. The bright spot is not usually seen in CVs during outburst, as the bright outbursting disc normally outshines it. Since these were calculated using the same zero phase for all lines, assumed from the He\,\textsc{ii} 4686 radial velocities, HJD$_0$\,=\,2456037.7551, the maps may be rotated due to the unknown phase shift between our zero phase and the true zero phase of the white dwarf. Note that the bright spot appears at slightly different phases in each line. The extended bright spot seen in H$\alpha$, overlaps in phase with the bright spots seen in both He\,\textsc{ii} 4686 and H$\beta$. Some evidence of the bright spot may also be seen in the Doppler maps of the He\,\textsc{i} lines, however, they are dominated by the brighter varying signal causing the band of increased emission in the right-hand quadrants.

We note that we do not detect the presence of spiral arms in our data. Spiral arms have been observed in a  number of dwarf novae during outburst (e.g.\ \citealt{2001LNP...573...45S,2002PASJ...54L...7B}), and are thought to be caused by the tidal affect of the donor on the large disc. As our observations may correspond to an early point in the evolution of the outburst, it is possible that spiral arms develop later in the outburst.

The H$\beta$ line profile reveals a strong feature in the high velocity wing of the blueshifted peak, centred at about $-$800\,km\,s$^{-1}$. The redshifted peak also extends beyond the range of the bright spot visible in the trailed spectrum. The origin of this feature is unclear, and it appears in the Doppler map as a ring of emission at higher velocity than most of the disc emission.

We detect no rotation between Doppler maps created using only the first and only the third orbit, further verifying our identification of the orbital period.

\begin{figure}
 \includegraphics[width=0.48\textwidth]{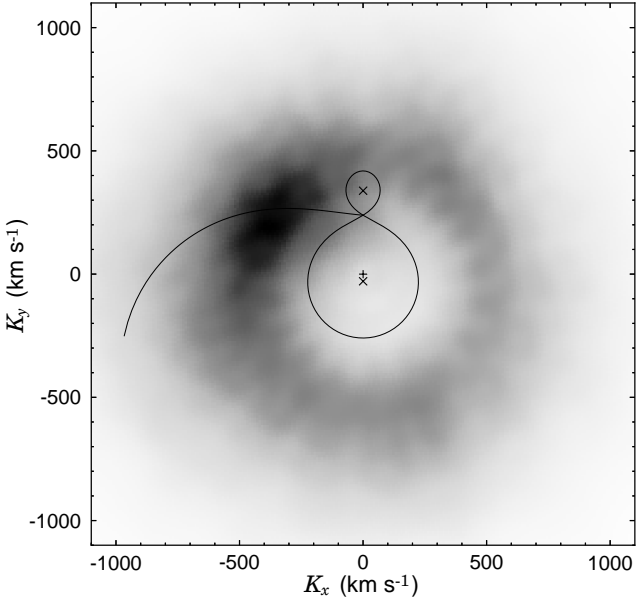}
 \caption{He\,\textsc{ii} 4686 Doppler map overplotted with Roche lobes and stream velocities of a $q$\,=\,0.086 binary. The map was rotated by applying a $-$0.08 phase shift compared to the maps in Fig. \ref{f:trail}. \label{f:stream}}
\end{figure}

We plot the Roche lobes and stream velocities for a $q$\,=\,0.086 binary together with the He\,\textsc{ii} 4686 Doppler map in Fig. \ref{f:stream}.
The velocity positions of the accretor, donor and centre of mass are also shown. Again, the map may be rotated about its origin due to the unknown phase shift between our assumed zero phase and the true zero phase.

\subsection{Grazing eclipse} \label{s:eclipse}

\begin{figure}
 \includegraphics[width=0.48\textwidth]{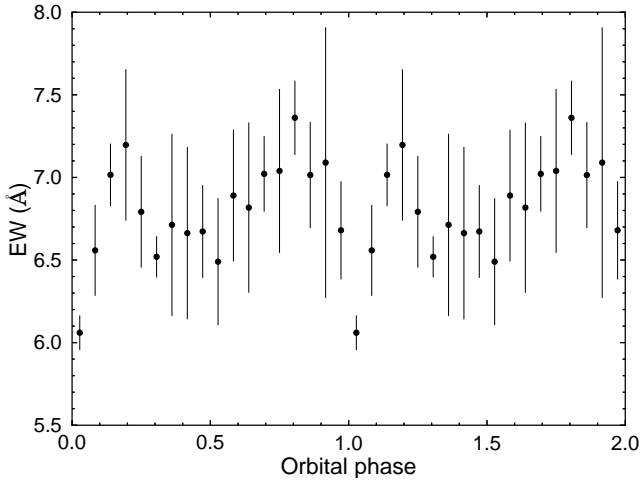}
 \caption{Binned, phase-folded He\,\textsc{ii} 4686 EW against orbital phase. The dip at about phase zero is attributed to a grazing eclipse of the accretion disc. Note that the EW axis has a narrow range, the depth of the eclipse is approximately 11 per cent.\label{f:ewlc}}
\end{figure}

The trailed spectra (Fig. \ref{f:trail}) show a slight reduction in the line flux at a phase of $\sim$1. To examine this further we construct a lightcurve of the He\,\textsc{ii} 4686 EW, shown in Fig. \ref{f:ewlc}. This reveals a clear dip, which we attribute to an eclipse of the outer disc.

This allows us to constrain the inclination of the system, as it must be large enough that the outer edge of the disc can be eclipsed. Since there is no eclipse of the inner disc or the accretor, we can also place an upper limit on its value. Using our derived value for $q$, the \citet{1983ApJ...268..368E} formula for the Roche lobe radius of the secondary, and the approximate tidal limit for the maximum radius of the accretion disc (e.g.\ \citealt{1995CAS....28.....W}), we find $63.8 \pm 1.0 ^{\circ} < i < 78.6 \pm 0.6 ^{\circ}$.

The outbursting disc is expected to be larger than the quiescent disc, significantly increasing the likelihood of a grazing eclipse during outburst. It is therefore likely that there will be no eclipse detectable in quiescence.


\section{Discussion}

The strength of the helium emission lines compared to the hydrogen lines in the average spectrum (Fig. \ref{f:qavspec} and Table \ref{t:ew}), highlights the unusual nature of this system. Whilst a detailed abundance analysis cannot be carried out with our current data, this is a strong indication of a much greater helium abundance than normally seen in CVs. Models of accretion discs and donors in CVs and AM CVns indicate that very little hydrogen is required to excite strong Balmer lines \citep{1982ApJ...257..672W,1991ApJ...366..535M,2009A&A...499..773N,2010MNRAS.401.1347N}, and the hydrogen abundance in SBSS\,1108+574 may be significantly lower than 10 per cent.

Our derived period, 55.3\,$\pm$\,0.8 minutes (26.03\,$\pm$\,0.38 cycles d$^{-1}$), is well below the CV period minimum, clearly indicating that the donor is significantly evolved, having been stripped of most of its hydrogen by mass-transfer or prior to the onset of mass-transfer. Our spectroscopic period favours the weak 55.36 minute (26.01\,$\pm$\,0.01 cycles d$^{-1}$) signal detected from the photometry and identified as the orbital period, over the 56.34 minute (25.56 cycles d$^{-1}$) signal identified as the superhump period. We are therefore confident that the candidate orbital period signal detected in our photometry and by \citet{2012arXiv1210.0678K}, is the correct orbital period. This confirms SBSS\,1108+574 as one of the shortest period CVs known, and places it well within the AM CVn period range.

There are three proposed formation channels for the AM CVn binaries, defined by the type of donor. The donor can be (1) a second, lower mass white dwarf \citep{1967AcA....17..287P,1972ApJ...175L..79F}, (2) a semi-degenerate helium star \citep{1986A&A...155...51S,1987ApJ...313..727I}, or (3) an evolved main-sequence star that has lost most of its hydrogen envelope. The latter is thought to form via the `evolved CV' channel \citep{2002ApJ...567L..49T,2003MNRAS.340.1214P}, but this formation channel has generally been considered to be unimportant in comparison to the double white dwarf and helium star channels. \citet{2003MNRAS.340.1214P}, however, argue that the evolved CV channel could contribute a significant fraction of the total AM CVn binary population. SBSS\,1108+574 has all the characteristics of an AM CVn progenitor in the evolved CV formation channel.

It must be noted that not all CVs with evolved donors reaching periods below the normal period minimum, will form AM CVn binaries as they are currently recognised (Yungelson et al., in preparation). Many systems will in fact reach their own period `bounce', and evolve back towards longer orbital periods, without depleting their hydrogen sufficiently to appear as AM CVn binaries. Current models suggest that the fraction of evolved CV channel AM CVn binaries may be lower than previously predicted (\citealt{2003MNRAS.340.1214P}; Yungelson et al., in preparation). It is clear, however, that systems like SBSS\,1108+574 and CSS1122$-$1110 \citep{2012MNRAS.425.2548B} fall between the standard definitions of AM CVn binaries and CVs.

The large mass ratio we derive, $q$\,=\,0.086\,$\pm$\,0.014, indicates a different evolution to the standard AM CVn binary population; for GP Com (P$_{\rmn{orb}}$\,=\,46.6\,minutes; \citealt{1999MNRAS.304..443M}), $q$\,=\,0.018 \citep{2007ApJ...666.1174R}, and for V396 Hya (P$_{\rmn{orb}}$\,=\,65.1\,minutes; \citealt{2001ApJ...552..679R}), $q$=0.013 \citep{2010htra.confE..10S}. The fact that the system shows outbursts indicates that the mass transfer rate is still relatively high, which is not expected for AM CVn binaries that have passed $P_{\rmn{min}}$ and evolved back to long periods (e.g.\ \citealt{2005ASPC..330...27N,2012A&A...544A..13K}). Compare our estimated accretion rate, $\dot M\sim10^{-10}\,\rmn{M}_\odot\,\mathrm{yr}^{-1}$ (for a $0.6\, \rmn{M}_\odot$ accretor), to the value \citet{2007ApJ...666.1174R} derived for GP Com, $\dot M < 3.6 \times 10^{-12}\,\rmn{M}_\odot\,\mathrm{yr}^{-1}$, and the value \citet{2009A&A...499..773N} estimated for V396 Hya, $\dot M\sim10^{-11}\,\rmn{M}_\odot\,\mathrm{yr}^{-1}$, again presenting a strong contrast between SBSS\,1108+574 and the long period AM CVn binaries. This indicates that SBSS\,1108+574 is still evolving towards shorter orbital periods, becoming increasingly helium-rich.


\section{Conclusion}

We have presented time resolved spectroscopy of the helium-rich dwarf nova SBSS\,1108+574 (SDSS\,J1111+5712), confirming the period detected photometrically during the 2012 April outburst. The system shows unusually strong helium emission in both outburst and quiescence, suggesting a high helium abundance. We measure the superhump period from our photometry as 56.34\,$\pm$\,0.18\,minutes, consistent with the result of \citet{2012arXiv1210.0678K}. The spectroscopic period is found to be 55.3\,$\pm$\,0.8\,minutes, significantly below the normal period minimum ($\sim$80\,minutes), confirming the system as an ultra-compact CV containing a highly evolved donor. The relatively high accretion rate, together with the large mass ratio, suggests that SBSS\,1108+574 is still evolving towards its period minimum.


\section*{Acknowledgements}

We thank the anonymous referee for useful comments and suggestions. PJC acknowledges the support of a Science and Technology Facilities Council (STFC) studentship. DS, TRM, BTG and EB acknowledge support from the STFC grant no. ST/F002599/1. T. Kupfer acknowledges support by the Netherlands Research School for Astronomy (NOVA). GHAR acknowledges an NWO-Rubicon grant. GN acknowledges an NWO-VIDI grant.

Based on observations obtained under programme GN-2012A-Q-54 at the Gemini Observatory, which is operated by the Association of Universities for
Research in Astronomy, Inc., under a cooperative agreement with the
NSF on behalf of the Gemini partnership: the National Science
Foundation (United States), the Science and Technology Facilities
Council (United Kingdom), the National Research Council (Canada),
CONICYT (Chile), the Australian Research Council (Australia),
Minist\'{e}rio da Ci\^{e}ncia, Tecnologia e Inova\c{c}\~{a}o (Brazil)
and Ministerio de Ciencia, Tecnolog\'{i}a e Innovaci\'{o}n Productiva
(Argentina).

Funding for the SDSS and SDSS-II has been provided by the Alfred P. Sloan Foundation, the Participating Institutions, the National Science Foundation, the U.S. Department of Energy, the National Aeronautics and Space Administration, the Japanese Monbukagakusho, the Max Planck Society, and the Higher Education Funding Council for England. The SDSS Web Site is http://www.sdss.org/.

Balmer/Lyman lines in the models were calculated with the
modified Stark broadening profiles of Tremblay \& Bergeron, ApJ 696,
1755, 2009, kindly made available by the authors.

\label{lastpage}

\end{document}